\documentclass[prb,aps,twocolumn,showpacs]{revtex4-2} 
\usepackage{amsmath,amssymb,amsthm}
\usepackage{graphicx}
\usepackage{subfigure}
\usepackage{amsmath}
\usepackage{amsfonts,amssymb}
\usepackage{bm}
\usepackage{float}
\usepackage{color}
\usepackage{sidecap}
\allowdisplaybreaks
\usepackage{soul}
\setstcolor{red}
\usepackage[normalem]{ulem}
\usepackage{hyperref}
\hypersetup{colorlinks=true,breaklinks,urlcolor=blue,linkcolor=blue,citecolor=blue}

\begin{document}

\title{Coexistence of topological Anderson insulator and multifractal critical phase in a non-Hermitian quasicrystal}

\author{Qi-Bo Zeng$^{1}$}
\email{zengqibo@cnu.edu.cn}
\author{Rong L\"u$^{2,3}$}
\affiliation{$^{1}$Department of Physics, Capital Normal University, Beijing 100048, China}
\affiliation{$^{2}$State Key Laboratory of Low-Dimensional Quantum Physics, Department of Physics, Tsinghua University, Beijing 100084, China}
\affiliation{$^{3}$Collaborative Innovation Center of Quantum Matter, Beijing 100084, China}

\begin{abstract}
The interplay of topology, disorder, and non-Hermiticity gives rise to phenomena beyond the conventional classification of quantum phases. We propose a one-dimensional non-Hermitian Su–Schrieffer–Heeger model with quasiperiodically modulated nonreciprocal intracell hopping. We show that quasiperiodic modulation can substantially enhance the topological regime and, remarkably, induce a non-Hermitian topological Anderson insulator (TAI) phase. Beyond the topological transition, increasing nonreciprocity drives a cascade of localization transitions in which all bulk eigenstates evolve from extended to multifractal critical and ultimately to localized states. Strikingly, the extended-to-critical transition coincides exactly with a real–complex spectral transition. We establish complete phase diagrams and derive exact analytical boundaries for both topological and localization transitions, uncovering an unanticipated coexistence of TAI and multifractal critical phases. Finally, we propose a feasible implementation in topolectrical circuits. Our results reveal a new paradigm for studying the cooperative effects of topology, quasiperiodicity, and non-Hermiticity.

\end{abstract}
\maketitle
\date{today}

\section{Introduction}
The discovery of topological phases has fundamentally reshaped our understanding of quantum matter, revealing boundary states protected by global topological invariants~\cite{Hasan2010RMP,Qi2011RMP,Armitage2018RMP}. While this framework was originally formulated for Hermitian systems, growing interest in open and dissipative platforms has propelled the development of non-Hermitian topology~\cite{Cao2015RMP,Konotop2016RMP,Ashida2020AP,Bergholtz2021RMP}. Non-Hermitian Hamiltonians naturally emerge in systems with gain and loss~\cite{Bender1998PRL,Bender2007RPP}, finite quasiparticle lifetimes~\cite{Kozii2017arxiv}, complex refractive indices in photonics~\cite{Musslimani2008PRL,Feng2017NatPho}, and engineered Laplacians in electrical circuits~\cite{Luo2018arxiv,Lee2018CommPhys,Sahin2025arxiv}. In contrast to Hermitian counterparts, such systems host intrinsically non-Hermitian topological structures, including exceptional points~\cite{Heiss2012JPA} and complex-energy braiding~\cite{Hu2021PRL,Wang2021Nature,Zhou2025PRB}. Remarkably, they also support phases without Hermitian analogs~\cite{Ashida2020AP,Bergholtz2021RMP}. A defining feature of nonreciprocal systems is the breakdown of conventional bulk–boundary correspondence due to the non-Hermitian skin effect (NHSE), where an extensive number of eigenstates accumulate at the boundaries~\cite{Yao2018PRL1,Yao2018PRL2}. The NHSE itself has a topological origin, intimately connected to point-gap topology under periodic boundary conditions~\cite{Borgnia2020PRL,Okuma2020PRL,Zhang2020PRL}.

Disorder, ubiquitous in realistic systems, generally destabilizes topological phases through Anderson localization~\cite{Anderson1958}. Yet disorder can also play a constructive role: Sufficiently strong randomness may induce topological transitions from trivial phases, leading to the topological Anderson insulator (TAI)~\cite{Li2009PRL,Groth2009PRL,Xing2011PRB,Jiang2009PRB,Zhang2012PRB,Song2012PRB,Atland2014PRL,Shem2014PRL,Zhang2019PRB,Hsu2020PRB,Liu2020PRL,Velury2021PRB,Zhang2021PRB,Lu2022AnndePhys,Ren2024PRL,Chen2024PRL,Ji2025arxiv}. The TAI phases and the topological properties based on the one-dimensional (1D) Su–Schrieffer–Heeger (SSH) model with quasiperiodic disorders have also been studied extensively~\cite{Liu2018PLA,Longhi2020OL,Tang2022PRA,Li2024PRR,Sircar2025PLA,Wang2026arxiv}. The interplay between disorder and non-Hermiticity further enriches localization and topological physics, producing unconventional phenomena that cannot be observed in the Hermitian systems~\cite{Hatano1996PRL,Shnerb1998PRL,Gong2018PRX,Longhi2019PRL,Longhi2019PRB,Jiang2019PRB,Zeng2020PRR,Liu2021PRB,Cai2021PRB,Chen2022PRB,Lin2022PRL,Sun2023PLA}. Extending the TAI paradigm to non-Hermitian systems has revealed a diverse landscape of disorder-driven topological transitions~\cite{Zhang2020SciChina,Liu2020CPB,Tang2020PRA,Tang2022PRA,Lin2022NatCom}. In parallel, quasiperiodic systems such as the André–Aubry–Harper model~\cite{Aubry1980,Harper1955} provide a deterministic route to localization transitions. With suitable quasiperiodic modulations or long-range hopping, quasicrystals can host mobility edges separating extended and localized states~\cite{Sarma1988PRL,Izrailev1999PRL,Ganeshan2015PRL,Deng2019PRL,Wang2020PRL}. The associated critical states may proliferate into an intermediate multifractal phase, distinguished from both extended and localized regimes by its spectral statistics and wave-function scaling properties~\cite{Geisel1991PRL,Jitomirskaya1999AnnMath,Halsey1986PRA,Mirlin2006PRL,Wang2021PRL,Wang2020PRL2,Xiao2021SciBull,Li2023npj}. Such critical phases are central to nonergodic dynamics, anomalous transport, and many-body localization phenomena~\cite{Pal2010PRB,Nandkishore2015ARCMP,Purkayastha2018PRB,Abanin2019RMP}. Topological states in such quasiperiodic lattices have also been extensively investigated~\cite{Lang2012PRL,Kraus2012PRL1,Kraus2012PRL2,Ganeshan2013PRL,Cai2013PRL,Wang2016PRB,Zeng2016PRB,Zeng2020PRB,Zeng2021PRB}. Despite these advances, a fundamental question remains unresolved: Can a disorder-induced TAI phase and a multifractal critical phase coexist within a unified non-Hermitian quasiperiodic framework? To date, these phenomena have largely been treated as distinct regimes, and the conditions enabling their simultaneous emergence remain unknown.

In this work, we resolve this question by introducing a one-dimensional non-Hermitian SSH model with quasiperiodically modulated nonreciprocal intracell hopping. We show that quasiperiodicity not only enhances the robustness of the topological phase but also drives a disorder-induced transition into a non-Hermitian TAI phase. More strikingly, increasing nonreciprocity induces a cascade of localization transitions, with all bulk eigenstates evolving from extended to multifractal critical and ultimately to localized regimes. The extended-to-critical transition coincides exactly with a real–complex spectral transition, establishing a direct correspondence between spectral feature and wave-function criticality. By deriving exact analytical phase boundaries, we construct the complete phase diagram and uncover a previously unexplored regime where the TAI and multifractal critical phases coexist. Finally, we propose a feasible implementation of our model in topolectrical circuits. Our work reveals the coexistence of a non-Hermitian TAI phase and a multifractal critical phase for all bulk states, where both the topological and the extended-critical-localized phase boundaries are analytically tractable. Hence our model provides an important platform to study the interplay of non-Hermiticity, topology, and quasiperiodic disorders.

The rest of the paper is organized as follows. In Sec.~\ref{Sec2}, we will first introduce the model Hamiltonian of the 1D non-Hermitian SSH model with quasiperiodically modulated nonreciprocal intracell hopping. In Sec.~\ref{Sec3}, we discuss the topological Anderson insulating phase and determine the phase boundaries. Then we will further investigate the extended-critical-localized phase transitions in Sec.~\ref{Sec4}. We also propose an experimental scheme for realizing the model in electric circuit in Sec.~\ref{Sec5}. Finally in Sec.~\ref{Sec6}, we will summarize our results. 

\section{Model Hamiltonian}\label{Sec2}
We consider a one-dimensional SSH chain with quasiperiodically modulated nonreciprocal intracell hopping, described by
\begin{equation}\label{H}
	H = H_{SSH} + H_\lambda,
\end{equation}
with 
\begin{equation}
	\begin{aligned}
		& H_{SSH} = \sum_j \left( v c_{j,A}^\dagger c_{j,B} + w c_{j,B}^\dagger c_{j+1,A} + h.c. \right), \\
		& H_\lambda = \sum_j \left( \lambda \cos \theta_j  c_{j,A}^\dagger c_{j,B} - \lambda \cos \theta_j c_{j,B}^\dagger c_{j,A} \right).
	\end{aligned}
\end{equation}
Here, $c_{j,A(B)}$ creates a spinless fermion on sublattice $A(B)$ of the $j$th unit cell. The parameters $v$ and $w$ denote the intracell and intercell hopping amplitudes, respectively, while $\lambda$ characterizes the strength of nonreciprocity. The modulation phase $\theta_j = 2\pi \alpha j + \phi$ introduces spatial inhomogeneity; for irrational $\alpha$, the nonreciprocity is quasiperiodic. Throughout this work we take $\alpha =(\sqrt{5}-1)/2$, $\phi=0$, and set $w=1$ as the energy unit. The lattice contains $N$ unit cells ($L=2N$ sites), and all parameters are real.

\begin{figure}[t]
	\includegraphics[width=3.4in]{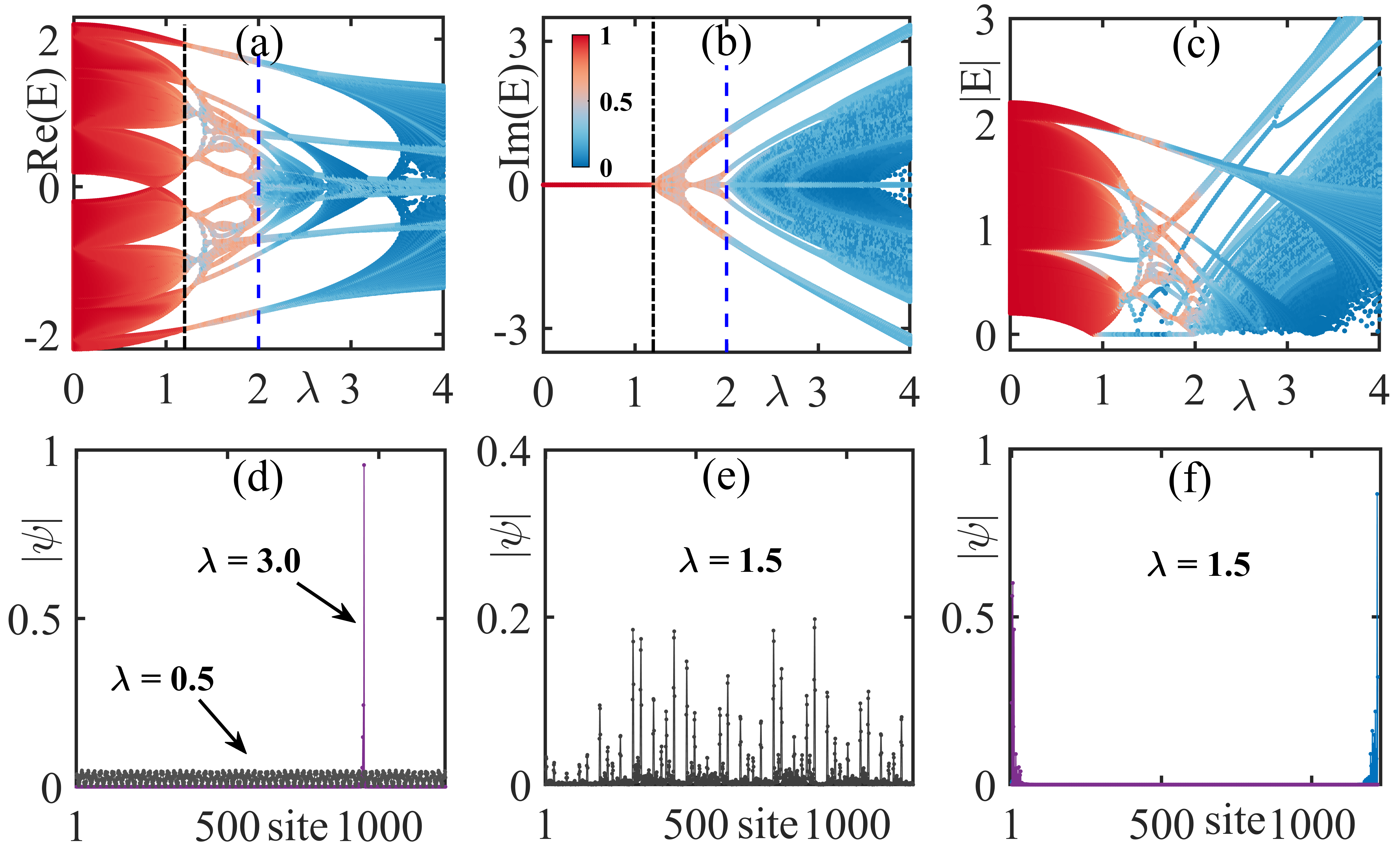}
	\caption{(Color online) (a) Real and (b) imaginary parts of the eigenenergies under periodic boundary conditions as a function of the nonreciprocity strength $\lambda$. The black and blue dashed lines mark the critical values $\lambda=v$ and $\lambda=2w$, separating the extended, multifractal critical, and localized regimes. The color scale indicates the fractal dimension of the corresponding eigenstates. (c) Energy spectrum $|E|$ under open boundary conditions. (d)–(e) Representative spatial profiles of eigenstates in the extended, critical, and localized phases, respectively. (f) Zero-energy edge modes in the topological Anderson insulating phase. Parameters: $v=1.2$, $w=1$, and system size $L=2N=1220$.}
	\label{fig1}
\end{figure}

The Hamiltonian takes a tridiagonal non-Hermitian form under open boundary conditions. For $|\lambda|<|v|$, it can be mapped to a Hermitian matrix via a similarity transformation $h=D^{-1}HD$ with diagonal $D$~\cite{Zeng2022PRB}(also see appendix for details). Consequently, $H$ is pseudo-Hermitian in the sense of Mostafazadeh~\cite{Mostafazadeh2002JMP}, and its spectrum is entirely real in this regime. Due to the quasiperiodicity of the nonreciprocal hopping, there will be no skin effect in this non-Hermitian model, and the spectra under open and periodic boundary conditions are consistent with each other~\cite{Zeng2020PRB2}. As illustrated in Fig.~\ref{fig1}(a) and \ref{fig1}(b), the eigenenergies remain real for $|\lambda|<|v|$ under periodic boundary conditions, while a real–complex spectral transition occurs at $|\lambda|=|v|$, beyond which the similarity transformation breaks down and complex eigenvalues emerge. 

To analyze the localization induced by quasiperiodic nonreciprocity, we formulate the transfer-matrix approach and compute the Lyapunov exponent. From the stationary Schrödinger equation $H | \psi \rangle = E | \psi \rangle$, with eigenstate $| \psi \rangle = \sum_j a_j | j,A \rangle + b_j | j, B \rangle$, yields
\begin{equation}\label{SchoEq}
	\left\{
	\begin{array}{ll}
		E a_j &=  (v + \lambda \cos \theta_j) b_j + w b_{j-1}, \\
		E b_j &=  (v - \lambda \cos \theta_j) a_j + w a_{j+1}.
	\end{array}
	\right.
\end{equation}
Introducing the two-component vector
\begin{equation}
	\Psi_j = \begin{pmatrix}
		a_j \\
		b_{j-1}
	\end{pmatrix}, \quad (b_{0}=0),
\end{equation}
the recursion relation can be written as
\begin{equation}
	\Psi_{j+1} = T_E(\theta_j) \Psi_j,
\end{equation}
with transfer matrix
\begin{equation}\label{TransMat}
	T_E(\theta_j)= \frac{1}{w(v+\lambda \cos \theta_j)} \begin{pmatrix}
		E^2-(v^2-\lambda^2 \cos^2 \theta_j) & -Ew\\
		Ew & -w^2
	\end{pmatrix}.
\end{equation}

The Lyapunov exponent is defined as 
\begin{equation}\label{Lyapunov}
	\gamma(E) = \lim_{N \rightarrow \infty } \frac{1}{N} \ln \left\| \prod_{j=1}^{N} T_E(\theta_j) \right\|, 
\end{equation}
where $N$ is the number of unit cells and $\left\| . \right\|$ denotes the matrix norm. The localization length is given by $\xi = \gamma^{-1} (E)$; thus $\gamma(E)>0$ signals localized state, while $\gamma(E)=0$ corresponds to extended or critical state.

To numerically characterize the localization features of the eigenstates, we can also compute the moments $I_q(n)=\sum_{j,\alpha} |\psi_{nR,j\alpha}|^{2q} \propto L^{-D_q(q-1)}$ with $\alpha = A$ and $B$, where $\psi_{nR,j\alpha}$ is the $j$th component on sublattice $\alpha$ of the $n$th right eigenstate $| \psi_{nR} \rangle$ and $D_q$ are the fractal dimensions. $q$ is a positive integer. If the eigenstates are localized (extended), we have $D_q \rightarrow 0$ $(1)$. While for the multifractal states, we have $0<D_q<1$~\cite{Evers2008RMP,Deng2019PRL}. Taking $q=2$ in $I_q$, we obtain the famous inverse participation ratio (IPR) as $\mathrm{IPR}(\psi_{nR}) = \sum_{j, \alpha} |\psi_{nR,j\alpha}|^4$ and fractal dimension as $D_2(\psi_{nR}) = - \lim_{L \rightarrow \infty} \left[ \ln \text{IPR}(\psi_{nR})/ \ln L \right]$, which are extensively used to characterize the localization properties. In the thermodynamic limit, $\mathrm{IPR}(\psi_{nR}) \rightarrow 0 $ for extended states and remains finite for localized states.

\section{Topological Anderson insulator}\label{Sec3}
For the clean SSH model ($\lambda=0$), the system is topologically nontrivial when $|v|<|w|$. Introducing quasiperiodic nonreciprocity fundamentally alters this criterion. Remarkably, we find that a topological phase can emerge in the parameter regime that is trivial in the clean limit, thereby realizing a non-Hermitian TAI. Figure~\ref{fig1}(c) shows the open-boundary spectrum $|E|$ as a function of $\lambda$ when $v=1.2>w$, where the clean system is topologically trivial. As $\lambda$ increases, zero-energy modes appear within a finite parameter window. Their spatial profiles [Fig.~\ref{fig1}(f)] reveal two states exponentially localized at opposite ends of the chain, confirming the onset of a nontrivial topological phase induced by quasiperiodic nonreciprocity.

The topological transition can be determined analytically from the divergence of the localization length of the zero-energy modes. Setting $E=0$ in Eq.~(\ref{SchoEq}), the two equations decouple and yield
\begin{equation}
	a_{j+1} = - \frac{v-\lambda \cos \theta_j}{w} a_j.
\end{equation}
Iterating from $j=1$ to $j=N$ gives
\begin{equation}
	a_{N+1} = (-1)^{N} \prod_{j=1}^{N} \frac{v-\lambda \cos \theta_j}{w} a_1.
\end{equation}
The Lyapunov exponent for the zero-energy mode is therefore
\begin{equation}
	\gamma(0) = \lim_{N \rightarrow \infty} \frac{1}{N} \sum_{j=1}^{N}  \ln \left| \frac{v-\lambda \cos \theta_j}{w} \right|.
\end{equation}
Since $\theta_j = 2\pi \alpha j + \phi$ with irrational $\alpha$, Weyl’s equidistribution theorem ensures that the sequence is uniformly distributed on $[0, 2\pi)$, The Lyapunov exponent reduces to the phase average~\cite{Weyl1916MathAnn,Choe993Proc,Longhi2019PRB}
\begin{equation}
	\gamma(0) = \frac{1}{2\pi} \int_{0}^{2\pi} \ln \left| \frac{v-\lambda \cos \theta}{w} \right| d \theta.
\end{equation}
Using Jensen’s formula, the integral can be evaluated exactly, yielding
\begin{equation}
	\gamma(0) = 	\left\{
	\begin{array}{ll}
		\ln \frac{\left| v \right| + \sqrt{v^2-\lambda^2}}{2 \left| w \right|}, \qquad  & \left| \lambda \right| < \left| v \right|; \\
		\ln \frac{\left| \lambda \right|}{2 \left| w \right| }, \qquad  & \left| \lambda \right| \geq \left| v \right|.
	\end{array}
	\right.
\end{equation}
The topological phase boundaries follow from $\gamma_0=0$,
\begin{equation}\label{TopoPhase}
	\begin{array}{ll}
		|v| + \sqrt{v^2-\lambda^2} = 2|w| \qquad  & \left| \lambda \right| < \left| v \right|; \\
	    |\lambda| = |2w|, \qquad  & \left| \lambda \right| \geq \left| v \right|.
	\end{array}
\end{equation}
These expressions provide exact analytical criteria for the disorder-induced topological phase transition.

\begin{figure}[t]
	\includegraphics[width=3.4in]{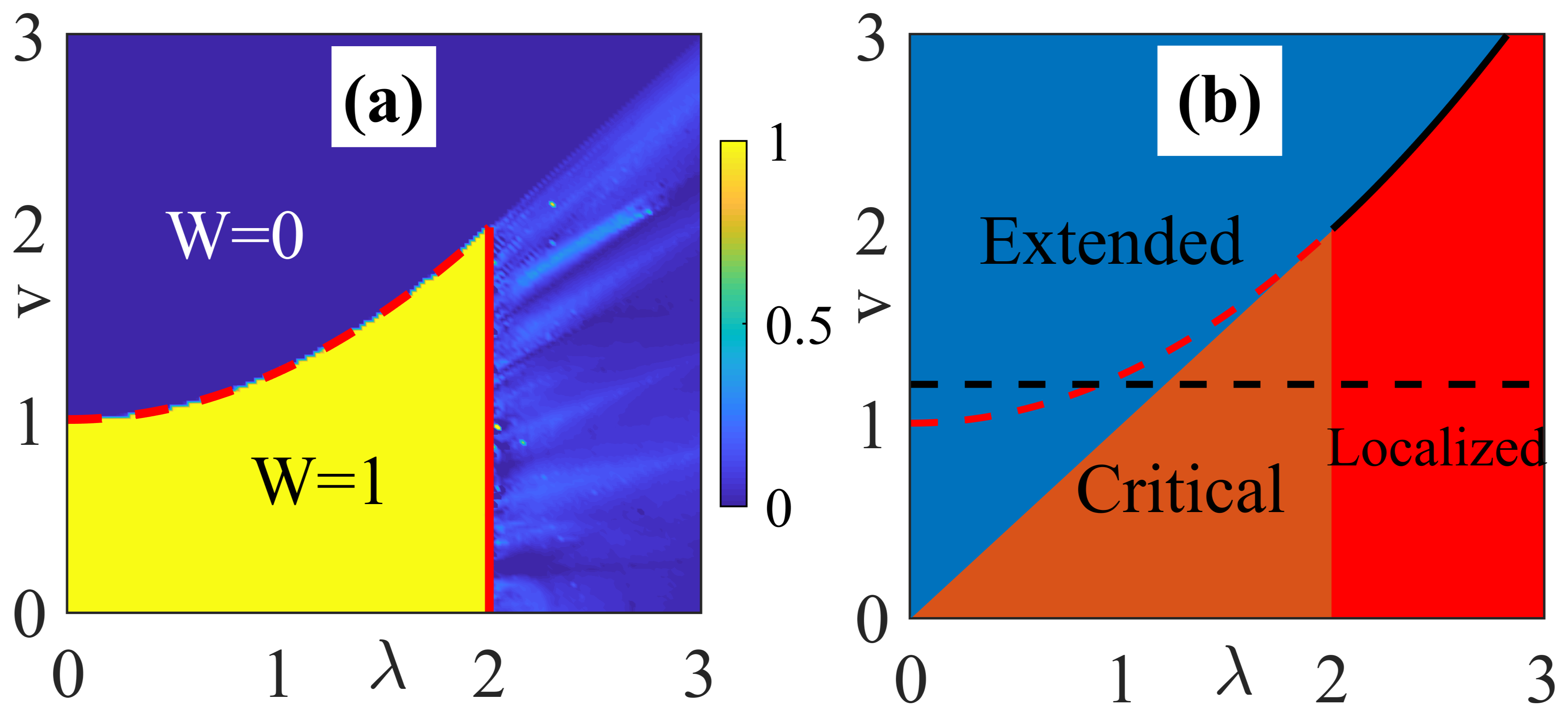}
	\caption{(Color online) (a) Phase diagram in the $v-\lambda$ plane showing the topologically nontrivial $(W=1)$ and trivial $(W=0)$ phases. The color scale represents the real-space winding number $W$. Red dashed and solid lines denote the analytical phase boundaries given by Eq.~(\ref{TopoPhase}). (b) Phase diagram of the extended, multifractal critical, and localized regimes determined from the Lyapunov exponent and fractal dimension. The red dashed line is the same as in (a). The black solid line is the phase boundary separating the extended and localized phase. The black dashed line indicates the cut at $v=1.2$.}
	\label{fig2}
\end{figure}

To further characterize the topological phase, we compute the real-space winding number under open boundary conditions~\cite{Song2019PRL}. The Hamiltonian preserves chiral symmetry, $SHS^{-1}=-H$, with $S=\sigma_z \otimes \mathcal{I}$. Let $| \psi^{\pm}_{nR} \rangle$ and $| \psi^{\pm}_{nL} \rangle$ denote the biorthonormal right and left eigenstates, which respectively satisfy the Schr\"odinger equations $H | \psi^{\pm}_{nR} \rangle = \pm E_n | \psi^{\pm}_{nR} \rangle$ and $H^\dagger | \psi^{\pm}_{nL} \rangle = \pm E_n^\star | \psi^{\pm}_{nL} \rangle$. Excluding edge modes, we define the open-boundary $Q$ matrix as $Q = \sum_n \left( | \psi^{+}_{nR} \rangle \langle \psi^{+}_{nL} | - | \psi^{-}_{nR} \rangle \langle \psi^{-}_{nL} | \right)$, and the winding number
\begin{equation}
	W = \frac{1}{2L^\prime} \mathrm{Tr^\prime} \left( SQ [Q, X] \right),
\end{equation} 
where $X$ is the coordinate operator and $\mathrm{Tr^\prime}$ denotes the trace over a central region of length $L^\prime$. In this work, we have taken the central region with $N/2 \leq n \leq 3N/2$ such that $L^\prime = N$ in the numeric implementation. We find $W=1$ in the topologically nontrivial regime and $W=0$ in the trivial regime.

The resulting phase diagram in the $v-\lambda$ plane is shown in Fig.~\ref{fig2}(a). We can see that in the region with $1<v<2$, the system is trivial when the disorder is weak. As $\lambda$ increases, the system will enter into the nontrivial phase with $W=1$, this is the topological Anderson insulating phase induced by the nonreciprocal quasiperiodic disorder. When $\lambda$ further increases, the system will become trivial again at $\lambda=2$. The boundaries separating the blue and yellow regions with different winding numbers, which are numerically obtained, agree precisely with the analytical boundaries in Eq.~(\ref{TopoPhase}), as indicated by the red dashed and solid lines in Fig.~\ref{fig2}(a), confirming the emergence of a disorder-induced topological phase.

\section{Extended-critical-localized phase transitions}\label{Sec4}
Quasiperiodic nonreciprocity also drives Anderson-type localization transitions. As shown in Figs.~\ref{fig1}(a) and \ref{fig1}(b), the spectrum separates into three distinct regimes characterized by the fractal dimension $\Gamma$ of the eigenstates. For $v=1.2$, all states are extended when $|\lambda|<1.2$, become multifractal critical in the intermediate region $1.2<|\lambda|<2$, and are fully localized for $|\lambda|>2$. Representative wave-function profiles are displayed in Figs.~\ref{fig1}(d) and \ref{fig1}(e), where the critical states exhibit clear multifractal spatial structures.

\begin{figure}[t]
	\includegraphics[width=3.4in]{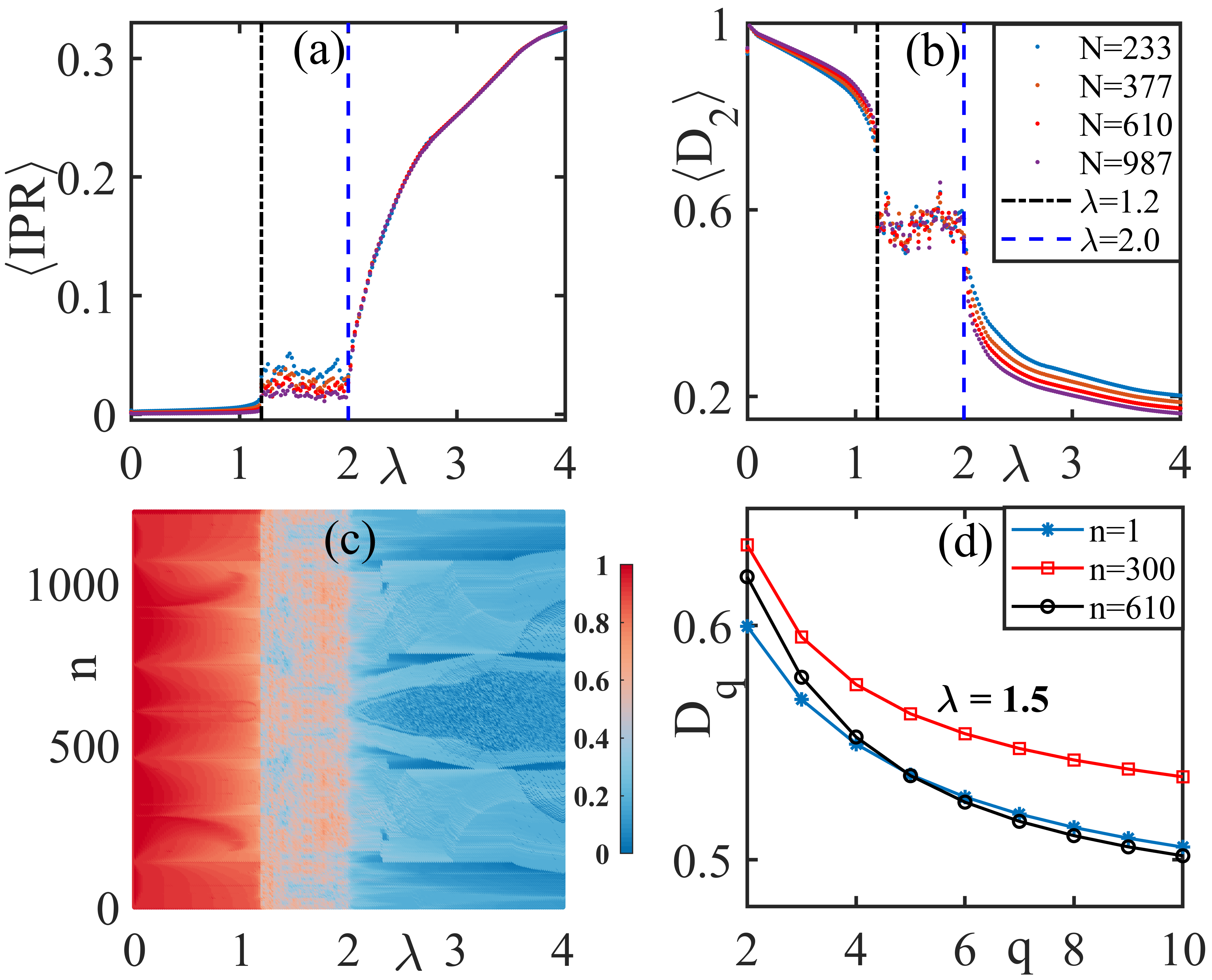}
	\caption{(Color online) (a) Average value of the inverse participation ratio $\langle IPR \rangle$ and (b) average value of the fractal dimension $\langle D_2 \rangle$ as functions of the nonreciprocity strength $\lambda$ for different system sizes under periodic boundary conditions. The black dot-dashed and blue dashed lines mark the analytical phase boundaries separating the extended–critical and critical–localized regimes, respectively. (c) The $D_2$ values of all the eigenstates of the system with $N=610$, which are indicated by the color scale. (d) Fractal dimensions $D_q$ for the $1$st, $300$th, and $610$th eigenstate at $\lambda=1.5$ in (c) as a function of $q$. Parameters: $v=1.2$ and $w=1$.}
	\label{fig3}
\end{figure}

To quantify the transition, we compute the average values of the inverse participation ratio and fractal dimension, which are respectively defined as $\langle IPR \rangle =\sum_n \left[\mathrm{IPR}(\psi_{nR})\right]/L$ and $\langle D_2 \rangle = \sum_n \left[D_2(\psi_{nR})\right]/L$. Figure~\ref{fig3}(a) and \ref{fig3}(b) show their dependence on $\lambda$ for different system sizes under periodic boundary conditions, such that the influences of topological edge modes are excluded. In the extended regime, $\langle IPR \rangle \rightarrow 0$ and $\langle D_2 \rangle \rightarrow 1$ as $L$ increases, while in the localized regime $\langle IPR \rangle$ remains finite and $\langle D_2 \rangle \rightarrow 0$. In contrast, both quantities remain taking intermediate values in the critical regime, and the $\langle D_2 \rangle$ behaves independently with the system size, consistent with multifractal scaling. Figure~\ref{fig3}(c) further plots the $D_2$ values of all eigenstates, where the color indicates the fractal dimensions of eigenstates. We can see that there is no mobility edges in the spectrum, and the boundaries indicated by the color scale at $\lambda=1.2$ and $2$ are quite clear, consistent with the results presented in Fig.~\ref{fig3}(a) and \ref{fig3}(b). For the multifractral critical phase, we further calculate the $D_q$ values of three representative eigenstates: $n=1$, $300$, and $610$, with $q$ varying from $2$ to $10$, as shown in Fig.~\ref{fig3}(d). We can see that the $D_q$ values are $q$ dependent, but always remain a finite value satisfying $0<D_q<1$, as expected for the multifractal states.

The localization transition can be determined analytically by using Avila’s global theory for quasiperiodic cocycles~\cite{Avila2015ActaMath,Zhou2023PRL}. For the case with $|\lambda|<|v|$, the transfer matrix defines an analytic cocycle $(\alpha, T_E(\theta))$. However, if $|\lambda| \geq |v|$, the transfer matrix in Eq.~(\ref{TransMat}) will become sigular due to the $(v+\lambda \cos \theta_j)$ term in the denominator and the cocycle $(\alpha, T_E(\theta))$ also becomes singluar. Since the Avila's global theory is developed to one-frequency analytic cocycles~\cite{Avila2015ActaMath}, we need to extend it to deal with the singular case here. To do so, we define the following new transfer matrix
 \begin{equation}
 	\widetilde{T}_E(\theta_j) = (v+\lambda \cos \theta_j) T_E(\theta_j).
 \end{equation}
 Then the $N$-step transfer matrix is $\widetilde{\mathcal{T}}_N= \left[ \prod_{j=1}^{N}(v+\lambda \cos \theta_j) \right] \mathcal{T}_N$, with $\mathcal{T}_N=\prod_{j=1}^{N} T_E(\theta_j)$, and we have
 \begin{equation}
 	\ln \left\| \widetilde{\mathcal{T}}_N \right\| = \sum_{j=1}^{N} \ln |v+\lambda \cos \theta_j| + \ln \left\| \mathcal{T}_N \right\|.
 \end{equation}
 According to the definition of Lyapunov exponent in Eq.~(\ref{Lyapunov}), the above formula leads to
 \begin{equation}
 	\widetilde{\gamma}(E) =  \gamma(E) + \lim_{N \rightarrow \infty } \frac{1}{N} \sum_{j=1}^{N} \ln |v+\lambda \cos \theta_j|.
 \end{equation}
 Since $\widetilde{T}_E(\theta_j)$ is analytic, $(\alpha, \widetilde{T}_E(\theta_j))$ defines an analytic Jacobi $GL(2,\mathcal{R})$ cocycle. The Avila's global theory can be extended to the non-singular Jacobi cocycles and the properties such as the acceleration-quantization still carries over to this case~\cite{Jitomirskaya2012ComMath}. Then we can use the Avila's theory to compute $\widetilde{\gamma}(E)$. Due to the ergodicity, the Lyapunov exponent admits a phase-averaged representation
 \begin{equation}
 \widetilde{\gamma}(E) = \lim_{N \rightarrow \infty} \frac{1}{2\pi N} \int_{0}^{2\pi} \ln \left\| \widetilde{T}_E(\theta_j) \right\| d \theta.
 \end{equation} 
By replacing $\theta$ with $\theta+iy$ with $y \in \mathcal{R}$, we get the complexified Lyapunov exponent and have
 \begin{equation}
 	\gamma(E,y) = \widetilde{\gamma}(E,y)  - \int_{0}^{2\pi} \ln |v+\lambda \cos (\theta + iy)| \frac{d \theta}{2\pi}.
 \end{equation}
 Setting $I(y)=\int_{0}^{2\pi} \ln |v+\lambda \cos (\theta + iy)| \frac{d \theta}{2\pi}$, one has
 \begin{equation}
 	 I(y) = \left\{
 	\begin{array}{ll}
 		\ln \frac{\left| v \right| + \sqrt{v^2-\lambda^2}}{2}, \qquad  & \left| \lambda \right| < \left| v \right|; \\
 		|y| +\ln \frac{|\lambda|}{2}, \qquad  & \left| \lambda \right| \geq \left| v \right|. 
 	\end{array}
 	\right.
 \end{equation} 
 Now since $\widetilde{T}_E(\theta_j)$ is analytic, Avila's global theory ensures that $\widetilde{\gamma}(E,y)$ is a continuous, convex, piecewise linear function with respect to $y$. For large $|y|$, the term $\lambda \cos (\theta + iy)$ behaves as 
  \begin{equation}
 	\lambda \cos (\theta + iy) \sim \frac{\lambda}{2} e^{|y|} e^{\pm i \theta}.
 \end{equation}
 In this limit, the transfer matrix becomes asymptotically triangular. Consequently, for the Lyapunov exponent $\widetilde{\gamma}(E,y)$, we have
 \begin{equation}
 	\widetilde{\gamma}(E,y) = 2|y| + \ln \left| \frac{\lambda^2}{4w} \right| + o(1),
 \end{equation}
 where $o(1) \rightarrow 0$ in the large $|y|$ limit. This leads to
 \begin{equation}
 	\gamma(E,y) = 2|y| + \ln \left| \frac{\lambda^2}{4w} \right| - I(y),
 \end{equation}
 and we have
 \begin{equation}
 	 	\gamma(E,y) =\left\{
 	\begin{array}{ll}
 		2|y| + \ln \frac{\left| v \right| - \sqrt{v^2-\lambda^2}}{2 \left| w \right|}, \qquad & \left| \lambda \right| < \left| v \right|; \\
 		|y| + \ln \left| \frac{\lambda}{2w} \right|, \qquad  & \left| \lambda \right| \geq \left| v \right|. 
 	\end{array}
 	\right.
 \end{equation}
The physical Lyapunov exponent corresponds to $\gamma(E,0)$.

From Avila's global theory, the physical Lyapunov exponent is obtained by matching the asymptotic behavior at large $|y|$, which gives
 \begin{equation}\label{gamma}
 	 	 	\gamma(E) =\left\{
 	\begin{array}{ll}
 		\text{max} \left\lbrace 0, \ln \frac{\left| v \right| - \sqrt{v^2-\lambda^2}}{2 \left| w \right|} \right\rbrace, \qquad  & \left| \lambda \right| < \left| v \right|; \\
 		\text{max} \left\lbrace 0, \ln \left| \frac{\lambda}{2w} \right| \right\rbrace, \qquad  & \left| \lambda \right| \geq \left| v \right|. 
 	\end{array}
 	\right.
 \end{equation}
 Remarkably, $\gamma(E)$ is independent of $E$. Therefore, all bulk states share the same localization behavior, and no mobility edge appears in the spectrum. If $\gamma(E)=0$, then the system is in the extended or critical phase; while if $\gamma(E)>0$, then the system is in the localized phase. The phase boundaries can be determined from the above formula by setting $\gamma(E)=0$, which leads to two critical lines
\begin{equation}
	|\lambda_{c1}| = 2|w| \sqrt{\left| \frac{v}{w}\right| -1}, \quad \text{and} \quad |\lambda_{c2}| = 2|w|,
\end{equation}
which coincides at $|v|=2|w|$. 

Note that if $|\lambda|<|v|<2|w|$, $\ln \frac{\left| v \right| - \sqrt{v^2-\lambda^2}}{2 \left| w \right|} < 0$, then according to Eq.~(\ref{gamma}), $\gamma(E)=0$ will always hold, implying that the system will in the extended or critical phase in this regime. On the other hand, if $|v| \leq |\lambda| < 2|w|$, then $\gamma(E)$ is determined by $\ln \left| \frac{\lambda}{2w} \right|$, and the critical disorder strength is $|\lambda_{c2}|=2|w|$. Thus, when $|\lambda| < 2|w|$, $\gamma(E)=0$, the system is in the extended or critical phase. If $|\lambda| > 2|w|$, then $\gamma(E) > 0$, and all eigenstates are localized. The distinction between extended and critical phases is controlled by the competition between $v$ and $\lambda$. When $|\lambda|<|v|$, the spectrum is absolutely continuous and all states are extended. For $|v|<|\lambda|<2|w|$, the spectrum becomes singular continuous~\cite{Avila2017InvenMath}, and all eigenstates are multifractal. Physically, the critical regime originates from incommensurate zeros in the effective hopping amplitudes when $|\lambda|>|v|$~\cite{Simon1989ComMath,Jitomirskaya2012ComMath}. So, for the region with $|v|<2|w|$, the system therefore exhibits three different phases
\begin{equation}
	\left\{
	\begin{array}{lll}
		\text{extended phase:} \quad  & 0 < |\lambda| < |v|,  \\
		\text{critical phase:} \quad & |v|  < |\lambda| < 2|w|, \\
		\text{localized phase:} \quad &|\lambda| > 2|w|.
	\end{array}
	\right.
\end{equation}
The resulting phase diagram is shown in Fig.~\ref{fig2}(b). Comparing with the topological phase diagram in Fig.~\ref{fig2}(a), we uncover a finite parameter region where the topologically nontrivial phase overlaps with the multifractal critical regime. Along the representative cut $v=1.2$, as indicated by the black dashed line in Fig.~\ref{fig2}(b), the system evolves as follows: Starting from a trivial extended phase at $\lambda=0$, increasing $\lambda$ first induces a TAI phase with extended bulk states; further increasing $\lambda$ drives the bulk into a multifractal critical phase while the zero-energy edge modes persist; finally, for $|\lambda|>2|w|$, the system becomes localized and topologically trivial. Notably, the extended–critical phase transition at $|\lambda|=|v|$ coincides exactly with the real–complex spectral transition, establishing a direct correspondence between the spectral property and wave-function criticality. Moreover, the critical window shrinks as $|v|$ increases and disappears entirely when $|v|>2|w|$. 

For the regime $|v|>2|w|$, the system exhibits the following two phases 
	\begin{equation}
		\left\{
		\begin{array}{ll}
			\text{extended phase:} \quad  & 0 < |\lambda| < 2|w| \sqrt{\left| \frac{v}{w}\right| -1},  \\
			\text{localized phase:} \quad &|\lambda| > 2|w| \sqrt{\left| \frac{v}{w}\right| -1},
		\end{array}
		\right.
	\end{equation}
which are separated by $|\lambda_{c1}|$. This phase boundary is plotted in Fig.~\ref{fig2}(b), as indicated by the black solid line separating the blue and red region.

Above we have analytically determined the phase boundaries for the extended-critical-localized phases in the SSH model nonreciprocal with quasiperiodic intracell hopping by extending the Avila's theory, the results are in good agreement with the numeric calculations.  We also reveal the coexistence of  multifractal critical phase and the nontrivial TAI phase induced by the nonreciprocal quasiperiodic modulation in the model, which have not been reported before.

\section{Experimental scheme in electric circuits}\label{Sec5}
\begin{figure}[t]
	\includegraphics[width=3.4in]{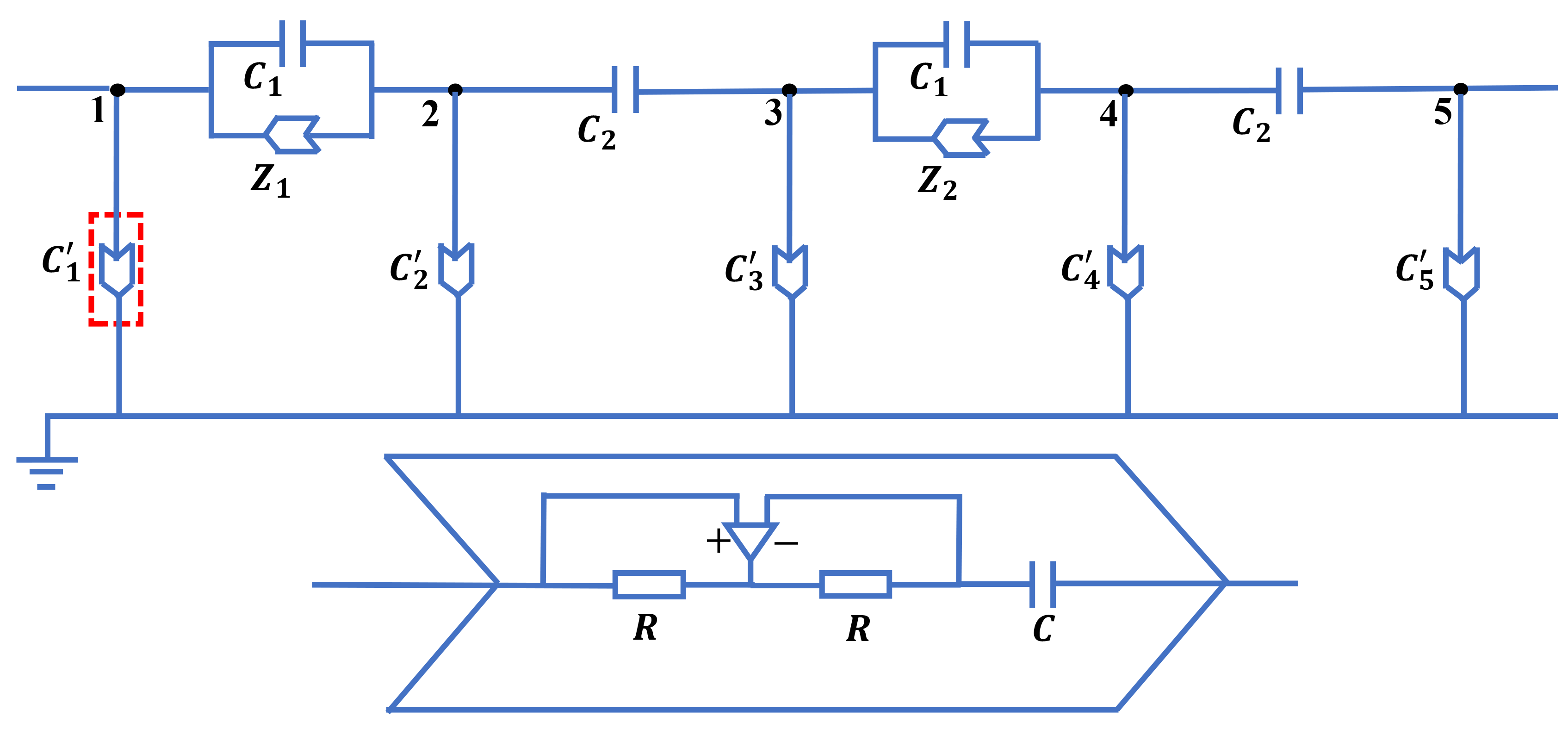}
	\caption{(Color online) Schematic of the topolectrical circuit realizing the non-Hermitian quasicrystal. The numbered black nodes represent lattice sites of the SSH chain. Capacitive couplings implement the Hermitian hopping terms, while the element highlighted in red denotes an INIC, which generates direction-dependent (nonreciprocal) intracell hopping. The effective impedances $Z_j$ and $Z_j^\prime$ are controlled by the orientation of the INIC. The lower panel illustrates the internal structure of the INIC, whose impedance depends on the direction of the current flow.}
	\label{fig4}
\end{figure}

Topolectrical circuits have emerged as a versatile platform for simulating topological band structures and non-Hermitian phenomena~\cite{Luo2018arxiv,Lee2018CommPhys,Sahin2025arxiv}. We propose a circuit implementation of the non-Hermitian quasicrystal, schematically shown in Fig.~\ref{fig4}. In the circuit network, nodes represent lattice sites of the SSH chain. The capacitors $C_1$ and $C_2$ realize the intra- and intercell hopping amplitudes, respectively. Nonreciprocal intracell hopping is implemented using negative-impedance converters with current inversion (INICs), whose direction-dependent impedance $Z_j$ generates the asymmetric hopping terms. By appropriately designing $Z_j$s and $Z_j^\prime$, the circuit Laplacian $\boldsymbol{J}$ defined through $\boldsymbol{I=JV}$ (with $\boldsymbol{I}$ and $\boldsymbol{V}$ the nodal current and voltage vectors), reproduces the effective Hamiltonian in Eq.~(\ref{H})(see the appendix). Topological and localization properties can be probed via impedance measurements. In particular, zero-energy edge modes manifest as pronounced peaks in the two-point impedance at the boundaries, while the admittance spectrum directly reveals the real–complex spectral transition and the localization regimes predicted in this work.

\section{Summary}\label{Sec6}
We have introduced and analytically characterized a non-Hermitian SSH quasicrystal with quasiperiodically modulated nonreciprocity. The interplay between non-Hermiticity and quasiperiodicity gives rise to two intertwined phenomena: a disorder-induced non-Hermitian TAI and a cascade of extended–critical–localized transitions. Most notably, we uncover a finite parameter regime where the TAI phase coexists with a multifractal critical bulk—an effect absent in previously studied disordered topological systems. We establish exact analytical phase boundaries and demonstrate that the extended–critical transition coincides precisely with the real–complex spectral transition, revealing a direct correspondence between spectral feature and wave-function criticality. The proposed topolectrical circuit implementation offers an experimentally accessible route to probe these multifractal topological states. More broadly, quasiperiodic nonreciprocity provides a general mechanism for engineering hybrid quantum phases in which topology, non-Hermiticity, and criticality are intrinsically intertwined.

\begin{acknowledgments}
	This work is supported by the National Natural Science Foundation of China (Grant No. 12204326). R.L. is supported by the Quantum Science and Technology-National Science and Technology Major Project (Grant No. 2021ZD0302100) and the "Gravitational Wave Detection" program (2023YFC2205800) funded by the Ministry of Science and Technology of the People's Republic of China.
\end{acknowledgments}

\setcounter{equation}{0}
\renewcommand*{\theequation}{A\arabic{equation}}
\begin{widetext}
\section*{Appendix}\label{Appendix}
In this Appendix, we will first give the proof for the pseudo-Hermicity of the model Hamiltonian and the conditions for the existence of real spectrum under open boundary conditions. Then we will provide more details on the experimental scheme for simulating the model Hamiltonian discussed in this work by using topolectric circuits. 

\subsection{The condition for a real spectrum}
Under open boundary conditions, the model Hamiltonian in Eq.~(\ref{H}) can be written as a tridiagonal matrix describing a one-dimensional nonreciprocal lattice. Defining the nearest-neighbor hopping amplitudes 
\begin{equation}
	t_{2j-1} = v+\lambda \cos \theta_j, \quad t_{2j-1}^\prime = v-\lambda \cos \theta_j, \quad t_{2j} = t_{2j}^\prime = w,
\end{equation}
where $t_n$ ($t^\prime_n$) is the backward (forward) hopping amplitude between the $n$th and the $(n+1)$th lattice sites. he Hamiltonian takes the form
\begin{equation}
	H = \left(
	\begin{array}{cccccc}
		0 & t_1 & 0 & 0 & \cdots & 0 \\
		t^\prime_1 & 0 & t_2 & 0 & \cdots & 0 \\
		0 & t^\prime_2 & 0 & t_3 & \cdots & 0 \\
		0 & 0 & t_3^\prime & 0 & t_4 & \cdots \\
		\vdots & \vdots & \vdots & \vdots & \ddots & \vdots \\
		0 & 0 & 0 & 0 & \cdots & 0
	\end{array}
	\right)
	=\left(
	\begin{array}{cccccc}
		0 & v+\lambda \cos \theta_1 & 0 & 0 & \cdots & 0 \\
		v-\lambda \cos \theta_1 & 0 & w & 0 & \cdots & 0 \\
		0 & w & 0 & v+\lambda \cos \theta_2 & \cdots & 0 \\
		0 & 0 & v-\lambda \cos \theta_2 & 0 & w & \cdots \\
		\vdots & \vdots & \vdots & \vdots & \ddots & \vdots \\
		0 & 0 & 0 & 0 & \cdots & 0
	\end{array}
	\right).
\end{equation}
All the elements in the matrix are real and we have $t_j \neq t^\prime _j$, implying that the matrix is non-Hermitian. 

Assume that $t_j t^\prime_j>0$, we construct a diagonal matrix
\begin{equation}\label{D}
	D = \left(
	\begin{array}{cccccc}
		d_1 & 0 & 0 & 0 & \cdots & 0 \\
		0 & d_2 & 0 & 0 & \cdots & 0 \\
		0 & 0 & d_3 & 0 & \cdots & 0 \\
		0 & 0 & 0 & d_4 & 0 & \cdots \\
		\vdots & \vdots & \vdots & \vdots & \ddots & \vdots \\
		0 & 0 & 0 & 0 & \cdots & d_L \\
	\end{array}
	\right),
\end{equation}
with
\begin{equation}\label{dj}
	d_n = \left\{
	\begin{array}{c}
		1, \qquad j = 1 \\
		\sqrt{\frac{t^\prime_{n-1} t^\prime_{n-2} \cdots t^\prime_1}{t_{n-1}t_{n-2} \cdots t_1}}, \qquad n = 2,3,\cdots,L
	\end{array}
	\right.
\end{equation}
Under the similarity transformation $h = D^{-1}HD$, the Hamiltonian becomes
\begin{equation}
	h = \left(
	\begin{array}{cccccc}
		0 & sgn(t_1)\sqrt{t_1 t^\prime_1} & 0 & 0 & \cdots & 0 \\
		sgn(t_1)\sqrt{t_1 t^\prime_1} & 0 & sgn(t_2)\sqrt{t_2 t^\prime_2} & 0 & \cdots & 0 \\
		0 & sgn(t_2)\sqrt{t_2 t^\prime_2} & 0 & sgn(t_3)\sqrt{t_3 t^\prime_3} & \cdots & 0 \\
		0 & 0 & sgn(t_3)\sqrt{t_3 t^\prime_3} & 0 & sgn(t_4)\sqrt{t_4 t^\prime_4} & \cdots \\
		\vdots & \vdots & \vdots & \vdots & \ddots & \vdots \\
		0 & 0 & 0  & 0 & \cdots & 0 \\
	\end{array}
	\right).
\end{equation}
Because condition $t_j t^\prime_j>0$ guarantees, all square roots are real and $h$ is Hermitian. Since similarity transformations preserve the spectrum, $H$ and $h$ share identical eigenvalues. Therefore, the spectrum of $H$ is entirely real whenever the condition $t_j t^\prime_j>0$ holds.

For the present model,
\begin{equation}
	t_{2j-1} t_{2j-1}'
	= (v+\lambda\cos\theta_j)(v-\lambda\cos\theta_j)
	= v^2-\lambda^2\cos^2\theta_j .
\end{equation}
The condition $t_j t^\prime_j>0$ for all $j$ requires
\begin{equation}
	v^2-\lambda^2\cos^2\theta_j > 0
	\quad \forall j,
\end{equation}
which is satisfied if and only if
\begin{equation}
	|\lambda| < |v|.
\end{equation}
Thus, the critical point for the real-to-complex spectral transition is
\begin{equation}
	|\lambda|=|v|.
\end{equation}
For $|\lambda|>|v|$, the similarity transformation breaks down because some $t_n$ and $t_n^\prime$ become negative, and complex eigenvalues necessarily emerge.

The above similarity transformation implies that the Hamiltonian is pseudo-Hermitian. Indeed,
\begin{equation}
	D^2 H^\dagger D^{-2} =D^2 (D h D^{-1})^\dagger  D^{-2}= D h D^{-1} = H,
\end{equation}
which yields $H^\dagger = \eta^{-1} H \eta$ with $\eta=D^2$. Thus, the Hamiltonian is pseudo-Hermitian. Therefore, $H$ is pseudo-Hermitian and admits a real spectrum in the parameter regime $|\lambda|<|v|$.

Notably, this argument does not depend on the specific functional form of $t_n$ and applies generally to one-dimensional nonreciprocal tridiagonal Hamiltonians~\cite{Zeng2022PRB}.

\subsection{Experimental scheme using topolectrical circuits}
The non-Hermitian quasicrystal proposed in the main text can be experimentally realized using topolectrical circuits, which have emerged as a versatile platform for simulating non-Hermitian band structures and topological phases. The circuit configuration is shown in Fig. 4 of the main text. The numbered black dots denote circuit nodes corresponding to lattice sites, where external voltage sources or impedance measurements can be applied.

The circuit consists of capacitors and INICs, which generate direction-dependent couplings and thereby emulate nonreciprocal hopping. The INIC structure (lower panel of Fig. 4) comprises two resistors, one capacitor, and an operational amplifier. Importantly, its effective impedance depends on the direction of current flow: It behaves as a negative capacitance for one orientation and as a positive capacitance for the opposite orientation, thus realizing asymmetric hopping amplitudes.

For a node $j$ in the circuit, let $I_j$ and $V_j$ denote the input current and voltage, respectively. Kirchhoff’s current law gives
\begin{equation}
	I_j = \sum_i Y_{ji} (V_j - V_i) + X_j V_j,
\end{equation}
where $Y_{ji}$ is the admittance between nodes $j$ and $i$, and $X_j$ denotes the shunt admittance at node $j$. Collecting all nodes, the voltage–current relation takes the matrix form
\begin{equation}
	\boldsymbol{I} = \boldsymbol{JV},
\end{equation}
where $\boldsymbol{J}$ is the circuit Laplacian. The circuit Laplacian can be used to model the tight-binding lattices Hamiltonian. 

For the electrical circuit shown in Fig.~\ref{fig4}, the Laplacian is
\begin{equation}
	\boldsymbol{J}=i\omega 
	\left(
	\begin{array}{cccccc}
		C_1+Z_1-C_1^\prime & -(C_1+Z_1) & 0 & 0 & 0 & \cdots \\
		-(C_1-Z_1) & C_1+C_2-Z_1-C_2^\prime & -C_2 & 0 & 0 & \cdots \\
		0 & -C_2 & C_1+C_2+Z_2-C_3^\prime & -(C_1+Z_2) & 0 & \cdots \\
		0 & 0 & -(C_1-Z_2) & C_1+C_2-Z_2-C_4^\prime & -C_2 & \cdots \\
		\vdots & \vdots & \vdots & \vdots & \ddots & \vdots \\
		0 & 0 & 0 & 0 & \cdots & C_1-Z_N-C_L^\prime
	\end{array}
	\right).
\end{equation}
This structure reproduces the tight-binding Hamiltonian under open boundary conditions, up to the overall factor $i \omega$.

To eliminate the diagonal (on-site) terms and obtain a purely off-diagonal SSH-type structure, we choose
\begin{equation}
	C_1^\prime=C_1 + Z_1, \qquad C_L^\prime=C_1 - Z_N, \qquad C_{2j}^\prime=C_1+C_2-Z_j \quad C_{2j-1}^\prime = C_1+C_2+Z_j \quad for \quad 1<j<N.
\end{equation}
The Laplacian then reduces to
\begin{equation}
	\boldsymbol{J}=i\omega 
	\left(
	\begin{array}{cccccc}
		0 & -(C_1+Z_1) & 0 & 0 & 0 & \cdots \\
		-(C_1-Z_1) & 0 & -C_2 & 0 & 0 & \cdots \\
		0 & -C_2 & 0 & -(C_1+Z_2) & 0 & \cdots \\
		0 & 0 & -(C_1-Z_2) & 0 & -C_2 & \cdots \\
		\vdots & \vdots & \vdots & \vdots & \ddots & \vdots \\
		0 & 0 & 0 & 0 & \cdots & 0
	\end{array}
	\right).
\end{equation}
Identifying
\begin{equation}
	C_1 \leftrightarrow v, \qquad C_2 \leftrightarrow w, \qquad Z_j=\lambda \cos(2\pi \alpha j + \phi),
\end{equation}
the circuit Laplacian becomes directly proportional to the non-Hermitian quasicrystal Hamiltonian discussed in the main text.

The admittance spectrum of the circuit corresponds to the energy spectrum of the Hamiltonian. In particular, zero-admittance resonances signal topological edge modes. The energy spectrum is obtained from the admittance spectrum of the circuit and can be used to detect the spectral properties reported in this work.

\end{widetext}

\end{document}